# Real-time Stress Measurements in Germanium Thin Film Electrodes during Electrochemical Lithiation/delithiation Cycling

Siva P.V. Nadimpalli,[a,*] Rajasekhar Tripuraneni,[a] Vijay A. Sethuraman[b]

[a] Department of Mechanical and Industrial Engineering, New Jersey Institute of Technology, Newark, New Jersey 07102, USA

[b] Department of Materials Engineering, Indian Institute of Science, CV Raman Avenue, Bangalore, Karnataka 560 012, India

[*] Corresponding Author, Email: siva.p.nadimpalli@njit.edu, Tel: +1 973 596 3678, Fax: +1 973 642 4282

## Abstract

An *in situ* study of stress evolution and mechanical behavior of germanium as a lithium-ion battery electrode material is presented. Thin films of germanium are cycled in a half-cell configuration with lithium metal foil as counter/reference electrode, with 1M $LiPF_6$ in ethylene carbonate, diethyl carbonate, dimethyl carbonate solution (1:1:1, wt. %) as electrolyte. Real-time stress evolution in the germanium thin-film electrodes during electrochemical lithiation/ delithiation is measured by monitoring the substrate curvature using the multi-beam optical sensing method. Upon lithiation *a*-Ge undergoes extensive plastic deformation, with a peak compressive stress reaching as high as -0.76±0.05 GPa (mean±standard deviation). The compressive stress decreases with lithium concentration reaching a value of approximately -0.3 GPa at the end of lithiation. Upon delithiation the stress quickly became tensile and follows a trend that mirrors the behavior on compressive side; the average peak tensile stress of the lithiated Ge samples was approximately 0.83 GPa. The peak tensile stress data along with the SEM analysis was used to estimate a lower bound fracture resistance of lithiated Ge, which is approximately 5.3 $J/m^2$. It was also observed that the lithiated Ge is rate sensitive, i.e., stress depends on how fast or slow the charging is carried out.

## 1. Introduction

Lithium-ion batteries are currently the primary choice as portable energy-storage devices for electronic devices, grid-storage, and automotive applications due to their high energy density among various existing battery chemistries[1]. However, the energy density of current lithium-ion batteries, which mainly use graphite as anode (372 mAh/g) and transition metal oxide as cathode (*e.g.*, $LiCoO_2$ with 140 mAh/g), is not sufficient to meet the future energy storage demands[2]. Extensive research is focused on finding alternative electrode materials for Li-ion batteries for future automotive applications that require higher energy densities. Silicon (Si), germanium (Ge), and their alloys are among the promising materials as anodes which have significantly higher charge capacities compared to that of graphite (*e.g.*, Si, and Ge have theoretical capacities of 3579 mAh/g [2] and 1625 mAh/g,[3] respectively).

Si has been studied extensively as a potential replacement for graphite due to its highest theoretical capacity among the alternatives to graphite as a lithium-ion battery anode. An analogous and isostructural group IV element Ge, however, did not receive similar attention primarily due to its higher cost. Ge has lower gravimetric capacity compared to Si but has a comparable volumetric capacity (7366 Ah/L for $Li_{15}Ge_4$ compared to 8334 Ah/L for $Li_{15}Si_4$)[4] and has much higher energy density compared to aluminum, tin, and graphite. Lithium diffusivity in germanium is two orders of magnitude higher than that in silicon, and germanium's electronic conductivity is higher than that of Si, [3, 5] which augment its rate capability. Ge has been shown to exhibit better cyclic performance [5] and superior fracture performance compared to Si.[6] The higher fracture resistance of lithiated Ge was attributed to the isotropic volume expansion of germanium as opposed to highly anisotropic volume expansion of Si when reacted with lithium[4, 6-7]. Hence, Ge can be a promising electrode for applications where durability, rate capability, and high energy density is important and cost may not be a primary issue, such as energy storage devices for space applications. Ge is abundant in nature and it is environmentally friendly material; the present cost of Ge is primarily driven by its low demand and hence can decrease if enough demand exists in future.

The higher capacity electrodes such as Si and Ge, however, pose some unique challenges. The volume expansion induced stresses in Si due to lithiation, for example, are as high as 1.7 GPa[8] and can influence the electrochemical processes. Sethuraman et al.[8-9] showed experimentally that the energy loss due to the stresses could be a significant portion of total battery losses and the electrode potentials are significantly affected by applied stresses. Similarly, a strong coupling exists between applied stresses and lithiation kinetics in Ge electrodes[10]. Meng Gu et al.[11] observed lithiation of a Ge nanowire in an *in situ* TEM study and demonstrated that stresses strongly influence the lithiation kinetics of Ge. They showed that the part of nanowire which was under tensile stress was favorable for lithiation reaction compared to the part that was under compressive stress.

Apart from affecting the kinetics of lithiation, stresses are the primary driving force for mechanical degradation of electrode materials[12] and effect the durability of electrodes. Some recent studies tried to synthesize microstructures based on Si-Ge nano structures to combine the superior rate capability of Ge with the higher capacity of Si[13]. However, without understanding the stress and mechanical property evolution in Ge due to lithiation, designing such innovative microstructures will not be efficient and may lead to premature failure. The critical role of mechanical stresses on the electrochemical performance and mechanical degradation of electrodes prompted several investigations; measurement of mechanical stresses and mechanical properties of the Si[8-9, 12], graphite[14], and Sn[15] electrodes are some examples. However, no study exists on the stress and mechanical property measurements of Ge during lithiation/delithiation cycling.

The objectives of this study are i) to carryout real-time stress measurement of Ge thin films during lithiation/delithiation cycling using the substrate curvature method, ii) to understand if the stress response of lithiated Ge is a function of charging rate, and iii) to estimate the lower bound fracture energy of lithiated Ge film To this end, Ge nano-films have been sputter deposited on a double side polished (DSP) fused silica wafers. The planar thin film geometry on the polished substrate not only allowed accurate stress measurements but is also ideal for conducting fundamental electrochemical and mechanical studies. The Ge thin film electrode was assembled in a half-cell configuration with a thin foil of lithium as reference/counter electrode. The electrodes were separated with a Celgard polymer separator. The Ge film was lithiated/delithiated galvanostatically and the stress evolution was measured. Similar to lithiated Si, Ge also exhibited extensive plastic deformation. The peak compressive stress of lithiated Ge was significantly lower than that of lithiated Si, but the tensile strength was almost similar which means that the losses associated with plastic dissipation will be less in Ge compared to Si. It was observed that lithiated Ge is rate sensitive, i.e., higher C rate will result in higher stress. Measured tensile stress was used in conjunction with the scanning electron microscopy (SEM) of the lithiated Ge film surface to estimate a lower bound fracture energy of the lithiated Ge film.

## 2. Experimental Procedures

### 2.1 Ge Thin Film Electrode and Electrochemical Cell Preparation

Electrodes were prepared by sputter depositing 5 nm of titanium (Ti), 200 nm of copper (Cu), and 100 nm of Ge films on 2-inch double-side polished fused silica substrates (525 μm thick, and

50.8 mm diameter). The Ge target was 99.999% pure with n-type doping. A schematic of different layers and their thickness is shown in Fig. 1a (inset). The role of the fused silica (or $SiO_2$ glass) substrate is to serve as an elastic substrate for the purposes of curvature measurements, and it does not participate in electrochemical reactions. The Ti layer acts as an adhesion layer between the $SiO_2$ substrate and the Cu film, which serves as the current collector. The films were deposited by RF-magnetron sputtering at a working pressure of less than 3 mTorr Ar while the base pressure before introducing the Ar gas was $4.4 \times 10^{-6}$ Torr. To minimize the thickness variations due to the relative position of gun and sample, the sample substrate was rotated during the deposition process. The sample substrate was cooled during the deposition process to minimize temperature rise and the associated residual stresses in the film. The residual stresses in Ge film was measured during sputter deposition using a sputter chamber equipped with substrate curvature measurement apparatus. Ge thin-films sputter deposited under these conditions are known to be amorphous.[5] In order to test this, Raman spectra were obtained on the sputter-deposited Ge thin films using a LabRAM scope (Horiba Scientific) with an Ar laser ($\lambda$=532 nm, 1 mW power) as the excitation source and are compared to that obtained on polycrystalline Ge pellets. The broad peak, in Fig. 1c, of sputtered Ge film confirms that the sputter deposited film is indeed amorphous. Planar thin film geometry without binder, shown in Fig. 1a, eliminates assumptions and complexities associated with the composite behavior; hence, this thin film configuration provides ideal conditions for measuring stress and other fundamental mechanical properties of lithiated germanium.

The electrochemical cell used in the experiments is illustrated schematically in Fig. 1a. The cell was assembled and tested in an argon-filled glove box (maintained at 25 °C and with less than 0.1 ppm of $O_2$ and $H_2O$). The Ge film was used as the working electrode and a lithium metal foil was used as the reference and counter electrode. The electrodes were separated by a Celgard polymer separator. 1 M LiPF$_6$ in 1:1:1 ratio (by wt.%) of ethylene carbonate (EC): diethyl carbonate (DEC): dimethyl carbonate (DMC) was used as the electrolyte. While the separator, lithium foil, and the Ge film stay submerged in the electrolyte, the top (reflecting) surface of the silica substrate remained above the electrolyte, which eliminates the optical complexities associated with the laser beam going through the electrolyte. A glass window ensured that the cell is sealed (and the composition of the electrolyte is preserved for the duration of the experiment).

## 2.2 *In situ* Measurement of Stress in Ge Thin Film during Lithiation

Stress evolution in the amorphous Ge (*a*-Ge) films during lithiation and delithiation was measured by monitoring the substrate curvature with the MOS setup (k-Space Associates, Dexter, MI) illustrated in Fig. 1a. The system consists of a laser source that generates a single focused beam, two etalons, whose rotation axes are arranged orthogonally to each other in order to generate a 2 x 2 array of beams, and a CCD camera to capture the reflected beam-array from the sample surface. The relative change in the distance between the laser spots on the CCD screen gives the sample curvature, $\kappa$, as

$$\kappa = \frac{\cos\phi}{2L}\left\{\frac{D-D_o}{D_o}\right\}\dots\dots\dots\dots\dots\dots\dots\dots\dots\text{Eq. 1}$$

Where $D$ is the distance between the laser spots and $D_o$ is its initial value. $\phi$ is the reflection angle of the beam, and $L$ is the optical path length from sample to the CCD camera, see Fig. 1b. The factor $\cos\phi/2L$, known as mirror constant, is specific to a setup and is obtained by calibrating the system in Fig.1a with a reference mirror of known curvature in the sample plane.

The 2 x 2 array of reflected spots enables curvature measurement in two orthogonal directions. The biaxial film stress is related to substrate curvature through the Stoney equation [16,17]

$$\sigma = \sigma_r + \frac{E_s t_s^2 \kappa}{6 t_f (1-v_s)} \quad \text{...........................Eq.2}$$

where $\sigma_r$ is residual stress in the film due to sputter deposition. $E_s$, $v_s$, and $t_s$ are Young's modulus, Poisson ratio, and thickness of the substrate, respectively. The $t_f$ is the film thickness of Ge which evolves continuously during lithiation and delithiation processes. Although there are no direct well controlled experiments on volume expansion measurements of lithiated Ge, measurement on Ge nanoparticles by transmission-electron microscopy studies revealed an approximate first order estimate; Liang et al. [6] reported a 260% increase in volume of spherical Ge nanoparticles upon full lithiation. A similar expansion was later confirmed in studies on Ge nanowires by Gu et al. [11]. Beaulieu et al.[18] conducted a controlled *in situ* thickness (and volume) evolution measurements with atomic force microscopy and showed that the thickness (and volume) of several anode materials (including Si which is isostructural to Ge) increases linearly with the state of charge (SOC). Hence, a linear variation of thickness with SOC during lithiation is assumed for Ge. The resulting equation for film-thickness as a function of SOC in Ge is given as

$$t_f = t_f^o (1 + 2.6z) \quad \text{......................................... Eq.3}$$

where $t_f^o$ is the initial film thickness, $z$ is SOC which varies between 0 and 1; $z=1$ corresponds to a capacity of 1625 mAh/g and a volumetric strain of 2.6.[6] It should be noted that the Eq. 2 is valid despite the changes in the thickness of the Ge film as per Eq.3 during electrochemical cycling; according to Freund and Suresh [16], the error caused by Eq.2 in the present case due to thickness changes of Ge is less than 0.04% which is negligible.

## 2.3 Electrochemical Measurements

Electrochemical experiments were carried out using a Solartron 1470 E potentiostat; cyclic voltammetry experiments were carried at a scan rate of 0.5 mV/s between 1.2 V and 0.05 V *vs*. Li/Li$^+$. For stress measurements, the Ge thin-film electrodes were lithiated galvanostatically at 5 µA/cm$^2$ with a lower cut-off potential of 0.05 V *vs*. Li/Li$^+$, delithiated at the same current density with an upper cut-off potential of 1.2 V *vs*. Li/Li$^+$. The cells were kept at open circuit for one minute between any two consecutive galvanostatic steps. Some of the cells were interrupted after one, three, and six lithiation/delithiation cycles; the samples were harvested, rinsed with dimethyl carbonate, and transported from inside the glove compartment to the SEM chamber in an argon-filled airtight container for SEM analysis. Care was taken to minimize the exposure of the samples to ambient atmosphere during this transfer. Some samples were subjected to multiple lithiation/delithiation cycles in order to measure the stress evolution in the Ge electrode beyond the first cycle and a few of them were lithiated to 0.005 V *vs*. Li/Li$^+$ to see if the phase changes (such as the formation of a crystalline Li$_{15}$Ge$_4$ phase reported to occur at potentials below 50 mV) have any effect on stress evolution. In order to understand the strain-rate sensitivity of the lithiated Ge, the electrodes were subjected to different current densities (0.5, 5, 50 µA/cm$^2$) at a given capacity during lithiation and corresponding stress response was recorded. Based on the lithium-diffusivity values, 1.5 X 10$^{-10}$ cm$^2$/s reported in the literature[5], the current densities are chosen such that they do not induce significant concentration/stress gradients in the film.

## 3. Results and Discussion

## 3.1 Stress and Potential Evolution in Ge Thin Films During lithiation/delithiation

Figure 2 shows the cyclic voltammogram (CV) data corresponding to the first 3 cycles obtained on the sputter deposited *a*-Ge thin-film electrodes. The sharp peak observed at *ca.* 600 mV in the first cycle, which diminished during the subsequent cycles, can be attributed to the formation of the solid-electrolyte-interphase (SEI) layer; besides this, a prominent peaks appear at 110 mV during first lithiation which corresponds to the lithium insertion into Ge. This peak shifted to *ca.* 125 mV in the subsequent cycles. A broad peak was observed during delithiation at *ca.* 430 mV which can be attributed to the extraction of lithium from Ge. These observations are consistent with the CVs reported on sputter deposited amorphous Ge films[3, 5].

Figures 3a and 3b show the potential and stress evolution, respectively, as a function of capacity in Ge films during galvanostatic lithiation/delithiation at 5μA/cm$^2$ (which corresponds to C/17.5 rate). The red and black curves in Fig. 3 indicate measurements from two different samples lithiated to different lower cut-off potential values: 5 mV and 50 mV, respectively. The open circuit potential of the cell prior to the initial lithiation was 2.8 V *vs.* Li/Li+, and as soon as the current was applied the potential dropped sharply to *ca.* 0.75 or 0.65 V *vs.* Li/Li$^+$; it decreases gradually to 0.05 V *vs.* Li/Li$^+$ or 0.005 V *vs.* Li/Li$^+$ thereafter. It can be noted, from Fig. 4a which depicts the potential evolution for three lithiation/delithiation cycles that the small plateau in potential at 0.75 V *vs.* Li/Li$^+$ was only observed in the first cycle. Hence, it can be attributed to solid electrolyte interphase (SEI) formation, which is consistent with the observation of Laforge *et al.*[5] and Graetz *et al.*[19]. Absence of any distinct constant potential regions in the potential curve suggests that there were no two-phase regions in the film during alloying (lithiation/delithiation) process and the addition of Li to amorphous Ge results in amorphous

Li$_x$Ge alloy, consistent with Laforge et al.[5]; although Laforge et al.[5] claim that there may be some domains of crystalized phases but the films consist of predominantly amorphous regions.

During lithiation, the substrate constrains the Ge film from in-plane (x-y plane) expansion, resulting in equi-biaxial compressive stress (indicated with a negative sign) in the film, Fig. 3b; the film can expand in the out-of-plane (z) direction only. Note that the stress curve starts at a value of -0.13 GPa which is the residual stress in Ge film due to deposition process. Initially, the compressive stress increases linearly with lithiation, which corresponds to the elastic response of the film, and becomes non-linear at approximately 110 mAh/g, indicating the onset of plastic flow of lithiated Ge. The stress reaches a peak value of -0.76 GPa at 135 mAh/g and decreases to -0.4 GPa at 600 mAh/g. Subsequent lithiation appears to result in only small changes in stress, gradually decreasing to about -0.23 or -0.3 GPa at a capacity of 1500 mAh/g. However, a distinct stress signature corresponding to the phase change associated with the amorphous Li$_x$Ge transforming in to a crystalline Li$_{15}$Ge$_4$ can be observed in the film lithitated to below 50 mV vs. Li/Li$^+$ potential. It appears, from inset showed in Fig. 3b, that the stress increases if lithiated beyond 50 mV which is an indication that the newly formed phases are relatively stronger (i.e., they can sustain slightly higher stress levels) than the amorphous phase that existed at 50 mV vs. Li/Li$^+$ potential. Upon delithiation (see Fig. 3b) the stress changes rapidly and becomes tensile with a small change in capacity of ~100 mAh/g, which represents elastic unloading-reloading of the film. Upon subsequent delithiation, the film begins to flow plastically in tension at a tensile stress of about 0.4 GPa. The film continues to flow plastically at a tensile stress of 0.5 GPa for the significant portion of delithiation and the stress begins to increase at a capacity of 600 mAh/g, which mirrors a similar decrease during lithiation between 150 and 600 mAh/g. This behavior was consistent both qualitatively and quantitatively among several samples. Although no studies exist currently on the deformation mechanism of lithiated Ge, the mechanisms in Ge could be similar to those observed in lithiated Si owing to similar crystal structure. Zhao et al.[20], based on density functional theory calculations, showed that the *a*-Si film could accommodate plastic deformation without further increase in the stress level, such as the stress behavior of lithiated Ge between 500 and 1000 mAh/g in Fig. 3b, due to the rearrangement of Si-Si bonds with Li-Si bonds. Hence, a similar bond rearrangement can be expected to happen in Ge film during lithiation. Since the Ge film continues to deform plastically, the measured stress history in Fig. 3b can be viewed as the evolving yield stress of lithiated Ge as a function of Li concentration.

The stress evolution behavior of lithiated Ge is qualitatively similar to that of lithiated Si[8, 12], although there are differences in capacity and stress magnitudes. For example, the peak compressive stress measured here, 0.76±0.05 GPa (average ± standard deviation), is significantly smaller than that of lithiated Si which is 1.5-1.75 GPa[8, 12] and 2 GPa[20]. It is interesting to note that although lithiated Ge films exhibit lower yield strength in compression, i.e., almost 1/3rd of lithiated Si yield strength, the tensile yield strength of lithiated Ge is almost similar to that of Si (e.g., 0.5 GPa for most part of delithiation). This means that the area enclosed by stress-capacity curve of lithiated Ge, which represents the energy loss due to plastic deformation, will be less compared to that of lithiated Si.

**3.2 Lower Bound Fracture Energy Estimate of Lithiated Ge film**

Figures 4b and 4c show that the yield stress values of lithiated Ge in 2$^{nd}$ and 3$^{rd}$ cycles match exactly with the values in the first cycle at any given capacity, which means that this Ge film did not undergo any mechanical damage such as cracking; an SEM analysis performed on the sample indeed confirmed this, see Fig. 5. This is in contrast to the lithiated Si which cracked within the first cycle when lithiated/delithiated under similar conditions[21]; this observation is consistent with earlier studies which reported that lithiated Ge is tougher than lithiated Si[6].

SEM images of the electrode before and after electrochemical cycling is shown in Fig. 5. Note that the surface after the first lithiation- delithiation cycle (Fig. 5b) and after 3 cycles (Fig. 5c) show no evidence of cracking. Although the film did not show cracking even after 6 cycles, Fig. 5d, the surface morphology has changed; features resembling anthill type of surface morphology can be seen in Fig. 5d. This could be due to nano-pore formation during repeated lithiation/delithiation of Ge electrode[22]; but, further studies are required to completely understand the mechanisms behind this surface evolution. However, it must be noted that the film is intact and did not develop any cracks after first three cycles. The lithiated Ge films in Figs. 5b and 5c sustained a peak tensile stress values of approximately 0.88 and 0.98 GPa, respectively, without developing any cracks. Hence, a fracture energy associate with a crack like defect under this peak stress value could represent a lower bound fracture energy of lithiated Ge film; in other words, one has to apply loads that exceed the estimated lower bound energy to cause fracture in the lithiated Ge film. In general, when the energy release rate, $G$ (represents an applied load), exceeds the fracture resistance of the film, $\Gamma$, cracks spread by channeling; this usually appears like a mud crack pattern[12]. The surface roughness of the substrate and other defects in the film possibly provide crack nucleation sites or flaws. By assuming the onset of cracking at the peak tensile stress, *i.e.*, 0.98 GPa (Fig. 3b), a lower-bound fracture energy per unit area, $\Gamma$, of lithiated Ge can be estimated.

Nakamura and Kamath[23] analyzed the problem of channel cracking in a film on an elastic substrate using a three-dimensional finite element method and showed that as soon as the flaw size reaches a value close to film thickness, $t_f$, the energy release rate becomes independent of crack length. Energy release rate for such a steady state channeling can be calculated by[24,25],

$$G = \frac{\pi}{2}\frac{(1-v_f^2)\,t_f\,\sigma_c^2}{E_f}g(\alpha,\beta)\,, \quad\text{Eq. 4}$$

where $v_f$, and $E_f$ are Poisson ratio and Young's modulus of the film, respectively. $\sigma_c$ is the critical stress where channel cracks propagate (assumed to be the peak stress in Fig. 3b), and the function $g$ depends on the elastic mismatch between the film and substrate through Dundur's parameters $\alpha$ and $\beta$. According to Xia and Hutchinson[24] and Beuth[25], the dependence of function $g$ on $\beta$ can be neglected; for the film/substrate properties of Table 1, $\alpha = -0.4$ and the function $g$ attains a value of 1.0. By substituting in Eq. 4, $t_f = 135$ nm, calculated according to Eq.3 at the peak tensile stress value of 0.98 GPa, and the Young's modulus value of the lithiated Ge (at 200 mAh/g) $E_f = 29$ GPa, average fracture energy of lithiated Ge is calculated to be 5.3 J/m². The Young's modulus value of lithiated Ge will change with lithium concentration; here the value at 200 mAh/g (corresponding to 0.98 GPa peak stress) is obtained by following the procedure similar to that of Sethuraman *et al.*[26], i.e., slope of red and green curves in Fig.4c at the beginning of second and third cycle lithiation (linear response corresponding to elastic response) represents the biaxial modulus of the film at 200 mAh/g. An ongoing project is focused on the measurement of modulus as a function of lithium concentration which will be reported later. As expected, the estimated average lower bound fracture energy of lithiated Ge, 5.3 J/m², is higher than the fracture energy of single crystal Ge[27] (111) which is about 1.75 J/m². In general, the ductile materials (such as lithiated Ge) have higher fracture resistance compared to brittle materials (such as pure single crystal Ge). However, it should be noted that the fracture energy obtained in this study is a first order estimate based on the assumptions behind stress measurements and volume expansion estimates. Nevertheless, the lower bound fracture energy of lithiated Si estimated with this approach[12] was in good agreement with the reported values from a different study[21]. Hence, the estimates provided here will be reasonably close to actual values and will be helpful in designing fracture resistant Ge nanostructures for lithium-ion battery electrodes.

**3.3 Strain-rate Sensitivity of Lithiated Ge Film**

Yield (or flow) stress of a rate sensitive material depends on how fast the material is strained (or deformed). Witvrouw and Spaepen[28] have showed that amorphous Si and Ge films exhibit rate dependent behavior; hence, it is possible that the lithiated Ge could exhibit a rate dependent deformation behavior. Strain is a direct function of, lithium concentration in Ge, applied current; hence, a higher applied current is equivalent to a higher strain-rate loading on the electrode material. Nadimpalli *et al.*[29] and Pharr *et al.*[30] showed that lithiated Si is a rate-sensitive material and its stress at any given SOC is a function of not only lithium concentration but also charging rate. Fig. 6 shows that the lithiated Ge also exhibits a rate-sensitive behavior. The Ge film was lithiated under galvanostatic conditions until the film starts to deform plastically and the applied current is varied in steps (*i.e.*, from a rate of C/175 to C/17.5 and from C/17.5 to C/1.75) within a small change of capacity. The corresponding potential and stress response is depicted in Fig. 6a and 6b. Note from Fig. 6c that the stress increases from -0.55 to -0.61 GPa when current density was changed from 0.5 to 5μA/cm$^2$ and from -0.61 to -0.66 GPa when current density increased from 5 to 50 μA/cm$^2$. This behavior was consistent among different samples. This suggests that depending on the rate of strain (or rate of charging) the stress levels could reach critical point leading to fracture which is consistent with recent reports on fracture of Ge pillars. For example, Lee *et al.*[7, 31] have conducted experiments on Si and Ge nano-pillars of different sizes under different lithiation/delithiation rates and demonstrated that higher charging rates promoted fracture of nano-pillars. Strain-rate sensitivity data along with the fracture energy estimates of lithiated Ge presented here will enable in ascertaining the critical charge rates that could cause particle cracking. Although mechanics based models for high capacity Li-ion battery electrodes such as [32] do consider strain-rate dependence of electrodes, the data in Fig.6c and the fracture data above will enable these models to predict the electrode behavior more accurately which helps not only in predicting fracture/failure of electrode particles but also in predicting the battery losses due to plastic deformation of electrode[8].

## 4. Conclusions

Germanium thin film electrodes were cycled against Li foil counter/reference electrode. Real-time stress evolution in planar a-Ge thin film electrodes is measured during galvanostatic lithiation/delithiation using substrate curvature technique. It was observed that upon lithiation *a*-Ge undergoes extensive plastic deformation, with a peak compressive stress as high as -0.76 GPa. The compressive stress decreased with lithium concentration reaching a value of -0.3 GPa at the end of lithiation. Upon delithiation the stress quickly became tensile and increased before reaching a plateau of 0.5 GPa until the capacity of the cell reached *ca.* 600 mAh/g, and started increasing with decrease in the capacity reached a peak tensile stress of 0.83 GPa. A distinct stress signature corresponding to the phase change associated with the amorphous $Li_xGe$ transforming into a crystalline $Li_{15}Ge_4$ can be observed in the film lithitated to below 50 mV *vs.* $Li/Li^+$ potential. It appears that the newly formed phases are relatively stronger (i.e., they can sustain slightly higher stress levels) than the amorphous phase that existed at 50 mV vs. $Li/Li^+$ potential. The stress evolution behavior of lithiated Ge is qualitatively similar to that of lithiated Si, although there are differences in capacity and stress magnitudes. For example, the peak compressive stress measured here, 0.76 GPa, is significantly smaller than that of lithiated Si which is 1.5-1.75 GPa and 2 GPa[20]. However, the tensile yield strength of lithiated Ge is almost similar to that of Si. This means that the area enclosed by stress-capacity curve of lithiated Ge, which is proportional to the energy loss due to plastic deformation, will be less compared to that of lithiated Si.

SEM analysis of the sample surfaces was carried out before and after electrochemical cycling. The surface after the first lithiation- delithiation cycle and after 3 cycles show no evidence of cracking. Although the film did not show cracking even after 6 cycles, the surface morphology has changed. This information along with peak tensile stress measurements were used to estimate a lower bound fracture energy of lithiated Ge film at low lithium concentration,which is approximately 5.3 $J/m^2$. Finally, it was observed that the lithiated Ge is rate sensitive, i.e., stress in the Ge film depends on the charge rate. Strain-rate sensitivity data along with the fracture energy estimates of lithiated Ge presented here will enable in ascertaining the critical charge rates that could cause particle cracking. The data reported here will enable the mechanics based models to predict the electrode behavior more accurately which helps not only in estimating the battery losses due to plastic deformation of electrodes but also in designing fracture/failure resistant Ge electrodes.

## 5. Acknowledgements

The authors gratefully acknowledge the financial support provided by New Jersey Institute of Technology through the faculty startup research grant.

## 6. References

1.    Tarascan, J.-M.; Armand, M., Issues and challenges facing rechargeable lithium batteries. *Nature* **2001,** *414*, 359.
2.    Uday Kasavajjula; Chunsheng Wang; Appleby, A. J., Nano- and bulk-silicon-based insertion anodes for lithium-ion secondary cells. *Journal of Power Sources* **2007,** *163*, 1003.

3.      Loïc Baggetto; Notten, P. H. L., Lithium-Ion (De)Insertion Reaction of Germanium Thin-Film Electrodes: An Electrochemical and In Situ XRD Study. *Journal of The Electrochemical Society* **2009,** *156* (3), A169.
4.      Xiao Hua Liu ; Yang Liu; Akihiro Kushima; Sulin Zhang; Ting Zhu; Ju Li; Huang, J. Y., In Situ TEM Experiments of Electrochemical Lithiation and Delithiation of Individual Nanostructures. *Advanced Energy Materials* **2012,** *2*, 722.
5.      B. Laforge; L. Levan-Jodin; R. Salot; Billardb, A., Study of Germanium as Electrode in Thin-Film Battery. *Journal of The Electrochemical Society* **2008,** *155* (2), A181.
6.      Wentao Liang; Hui Yang; Feifei Fan; Yang Liu; Xiao Hua Liu; Jian Yu Huang; Ting Zhu; Zhang, S., Tough Germanium Nanoparticles under Electrochemical Cycling. *ACS Nano* **2013,** *7* (4), 3427.
7.      Seok Woo Lee; Ill Ryu; William D. Nix; Cui, Y., Fracture of crystalline germanium during electrochemical lithium insersion. *Extreme Mechanics Letters* **2015**.
8.      Vijay A Sethuraman; M.J. Chon; M. Shimshak; V. Srinivasan; Guduru, P. R., In Situ Measurements of Stress Evolution in Silicon Thin Films during Electrochemical Lithiation and Delithiation. *Journal of Power Sources* **2010,** *195* (15), 5062.
9.      Vijay A Sethuraman; V. Srinivasan; A.F Bower; Guduru, P. R., In Situ Measurements of Stress-Potential Coupling in Lithiated Silicon. *Journal of The Electrochemical Society* **2010,** *157* (11), A1253.
10.     Hui Yang; Wentao Liang; Xu Guo; Chong-Min Wang; Zhang, S., Strong kinetics-stress coupling in lithiation of Si and Ge anodes. *Extreme Mechanics Letters* **2015**.
11.     Meng Gu; Hui Yang; Daniel E. Perea; Ji-Guang Zhang; Sulin Zhang; Wang, C.-M., Bending-Induced symmetry breaking of lithiation in germanium nanowires. *Nano Letters* **2014,** *14*, 4622.
12.     S.P.V. Nadimpalli; Vijay A. Sethuraman; G. Bucci; V. Srinivasan; A.F. Bower; Guduru, P. R., On Plastic Deformation and Fracture in Si Films during Electrochemical Lithiation/delithiation Cycling. *Journal of the Electrochemical Society* **2013,** *160* (10), A1885.
13.     Taeseup Song; Huanyu Cheng; Heechae Choi; Jin-Hyon Lee; Hyungkyu Han; Dong Hyun Lee; Dong Su Yoo; Moon-Seok Kwon; Jae-Man Choi; Seok Gwang Doo; Hyuk Chang; JIan Liang Xiao; Yonggang Huang; Won Park; Yong-Chae Chung; Hansu Kim; John A. Rogers; Paik, U., Si/Ge Double-layered Nanotube Array as a Lithium Ion Battery Anode. *ACS Nano* **2012,** *6* (1), 303.
14.     Amartya Mukhopadhyaya; Anton Tokranova; Xingcheng Xiao; Sheldona, B. W., Stress development due to surface processes in graphite electrodes for Li-ion batteries: A first report. *Journal of Power Sources* **2012,** *66*, 28.
15.     Hadi Tavassol; Michael W. Cason; Ralph G. Nuzzo; Gewirth, A. A., Influence of Oxides on the Stress Evolution and Reversibility during SnOx Conversion and Li-Sn Alloying Reactions. *Advanced Energy Materials* **2015,** *5* (1), 1400317.
16.     L. B. Freund; Suresh, S., Thin Film Materials: Stress, Defect Formation and Surface Evolution. **2004**.
17.     Stoney, G. G., *Proc. R. Soc. (Lond.)* **1909,** *172*, A82.
18.     L. Y. Beaulieu; T. D. Hatchard; A. Bonakdarpour; M. D. Fleischauer; Dahn, J. R., Reaction of Li with Alloy Thin Films Studied by In Situ AFM. *Journal of The Electrochemical Society* **2003,** *150* (11), A1457.

19. J. Graetz; C. C. Ahn; R. Yazami; Fultz, B., Nanocrystalline and Thin Film Germanium Electrodes with High Lithium Capacity and High Rate Capabilities. *Journal of The Electrochemical Society* **2004,** *151* (5), A698.
20. K. Zhao; G.A. Tristsaris; M. Pharr; W.L. Wang; O. Okeke; Z. Suo; J.J. Vlassak; Kaxiras, E., Reactive Flow in Silicon Electrodes Assisted by the Insertion of Lithium. *Nano Letters* **2012,** *12*, 4397.
21. Pharr, M.; Suo, Z.; Vlassak, J. J., Measurements of the Fracture Energy of Lithiated Silicon Electrodes of Li-Ion Batteries. *Nano Letters* **2013,** *13* (11), 5570-5577.
22. Xiao Hua Liu; Shan Huang; S. Tom Picraux; Ju Li; Ting Zhu; Huang, J. Y., Reversible Nanopore Formation in Ge Nanowires during LithiationDelithiation Cycling: An In Situ Transmission Electron Microscopy Study. *Nano Letters* **2011,** *11*, 3991.
23. T. Nakumara; Kamath, S. M., *Mechanics of Materials* **1992,** *13*, 67.
24. Xia, Z. C.; Hutchinson, J. W., *J. Mech. Physics of Solids* **2000,** *48*, 1107.
25. Jr, J. L. B., *Int. J. Solids and Structures* **1992,** *29*, 1657.
26. Sethuraman, V. A.; Chon, M. J.; Shimshak, M.; Winkle, N. V.; Guduru, P. R., In situ measurement of biaxial modulus of Si anode for Li-ion batteries. *Electrochemistry Communications* **2010,** *12*, 1614.
27. Pharr, G. M., Measurement of mechanical properties by ultra-low load indentation. *Materials Science and Engineering* **1998,** *A253*, 151.
28. Witvrouw, A.; Spaepen, F., Viscosity and elastic constants of amorphous Si and Ge. *Journal of Applied Physics* **1993,** *74* (12), 7154-7161.
29. Nadimpalli, S. P. V.; Buchovecky, E.; Sethuraman, V. A.; Shenoy, V. B.; Bower, A. F.; Guduru, P. R., Mechanical Characterization of Lithiated Silicon and the Solid Electrolyte Interpahse. *SES 49th Annual Technical Meeting, Georgia Tech, Oct 10-12* **2012**.
30. Pharr, M.; Suo, Z.; Vlassak, J. J., Variation of stress with charging rate due to strain-rate sensitivity of silicon electrodes of Li-ion batteries. *Journal of Power Sources* **2014,** *270*, 569.
31. S.W.Lee; McDowell, M. T.; Berla, L. A.; Nix, W. D.; Cui, Y., Fracture of crystalline silicon nanopillars during electrochemical lithium insertion. *Proceedings of National Society of Sciences* **2012,** *109* (11), 4080.
32. Bower, A. F.; Guduru, P. R.; Sethuraman, V. A., A Finite Strain Model of Stress, Diffusion, Plastic Flow and Electrochemical Reactions in a Lithium-ion Half-cell. *J. Mech. Phys. Solids* **2011,** *59* (4), 804.

# Figures

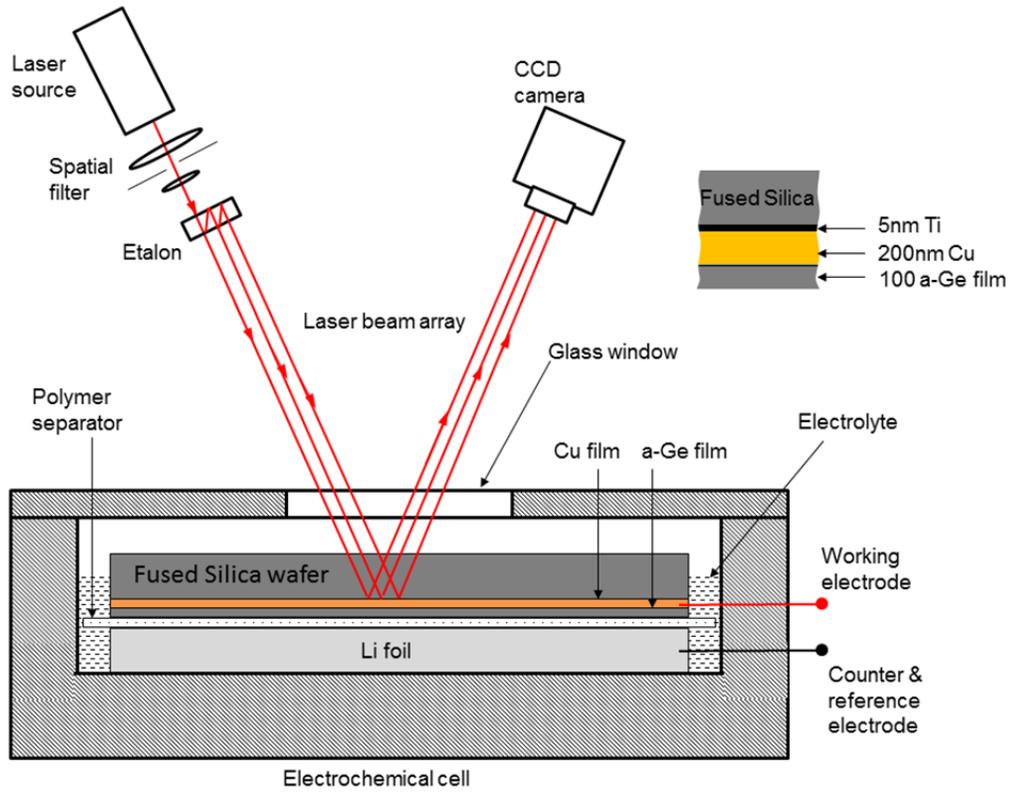

(a)

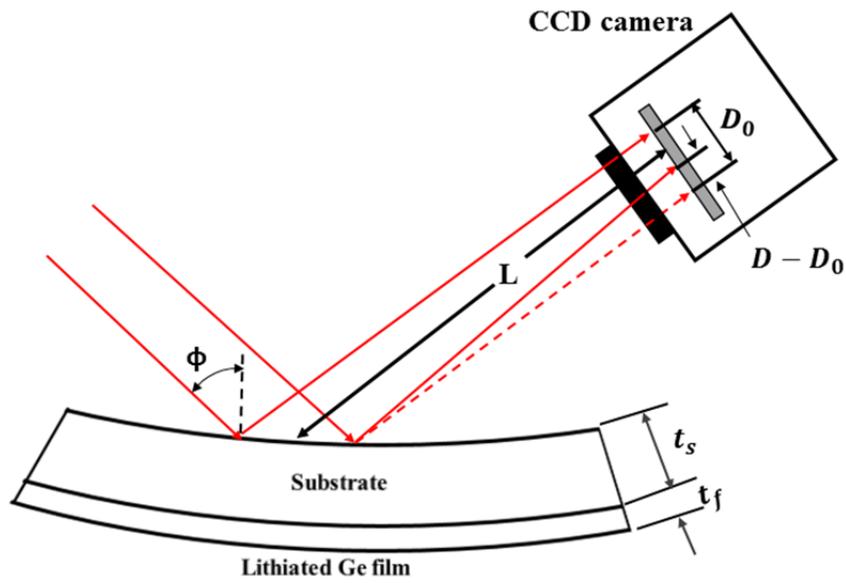

(b)

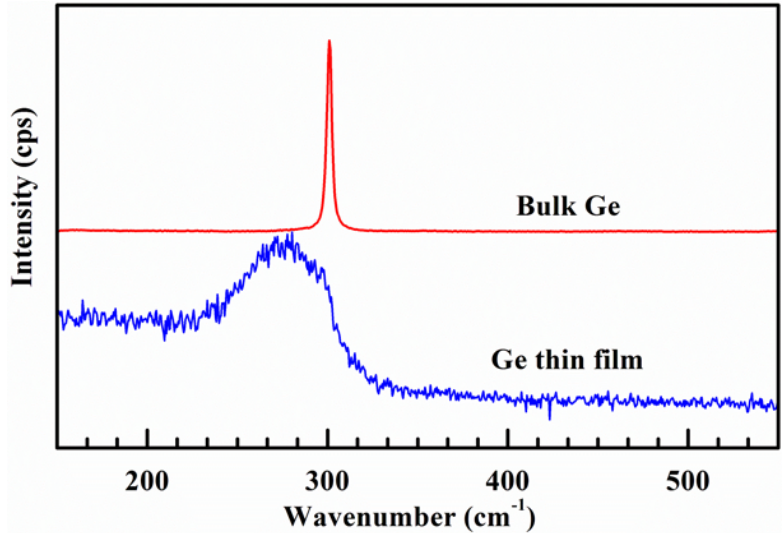

(c)

Figure1: (a) Schematic of experimental setup and electrochemical cell; the inset shows details of thin film layers and (b) shows the definition of various parameters used in Eq.1 and Eq.2 to measure curvature and stress. (c) Comparison of Raman spectra of sputter-deposited Ge thin films and polycrystalline (or Bulk) Ge. The broad peak of sputter-deposited film indicates that it is amorphous.

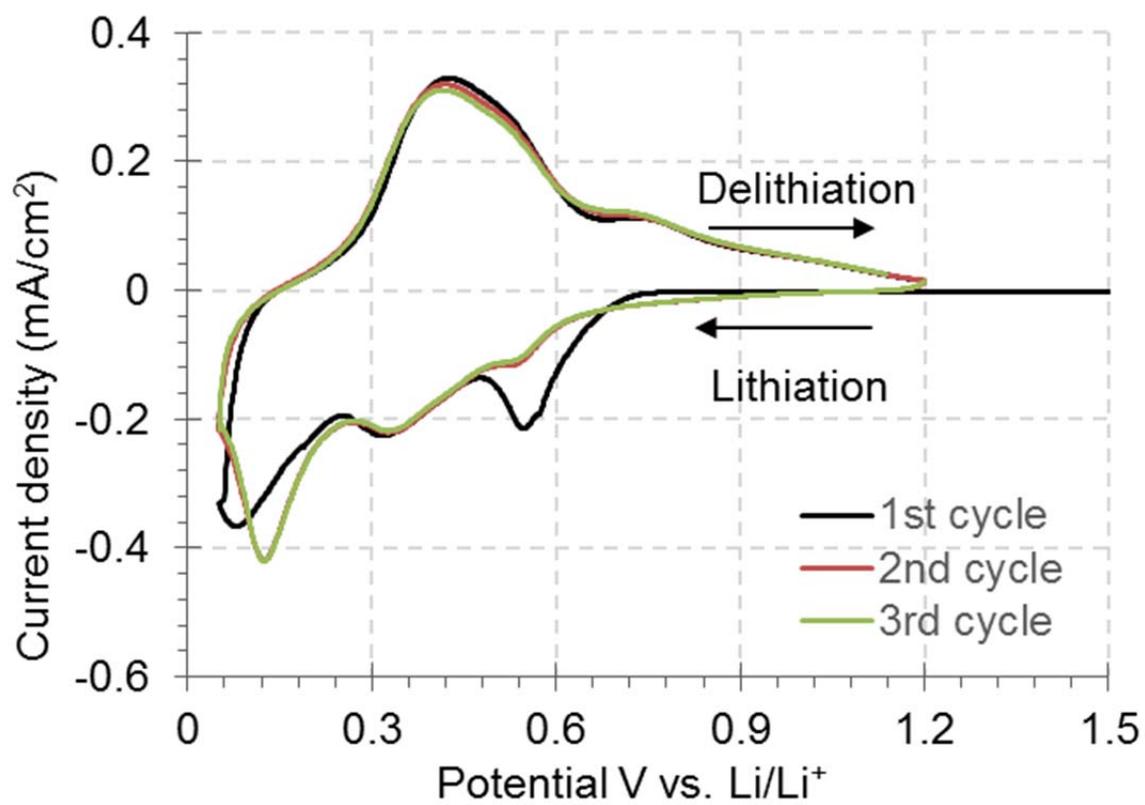

Figure 2: First three cycles from a cyclic voltammogram on a thin-film Ge electrode in 1M $LiPF_6$ in EC:DEC:DMC (1:1:1, wt.%) at a scan rate of 0.5 mV/s.

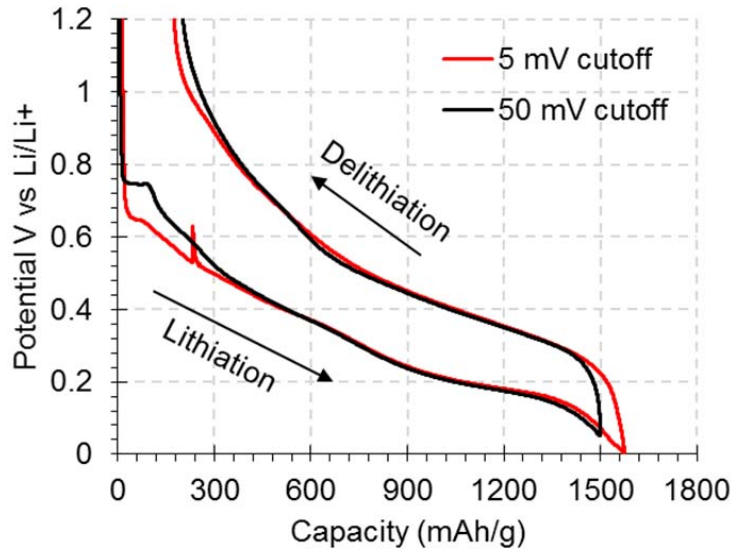
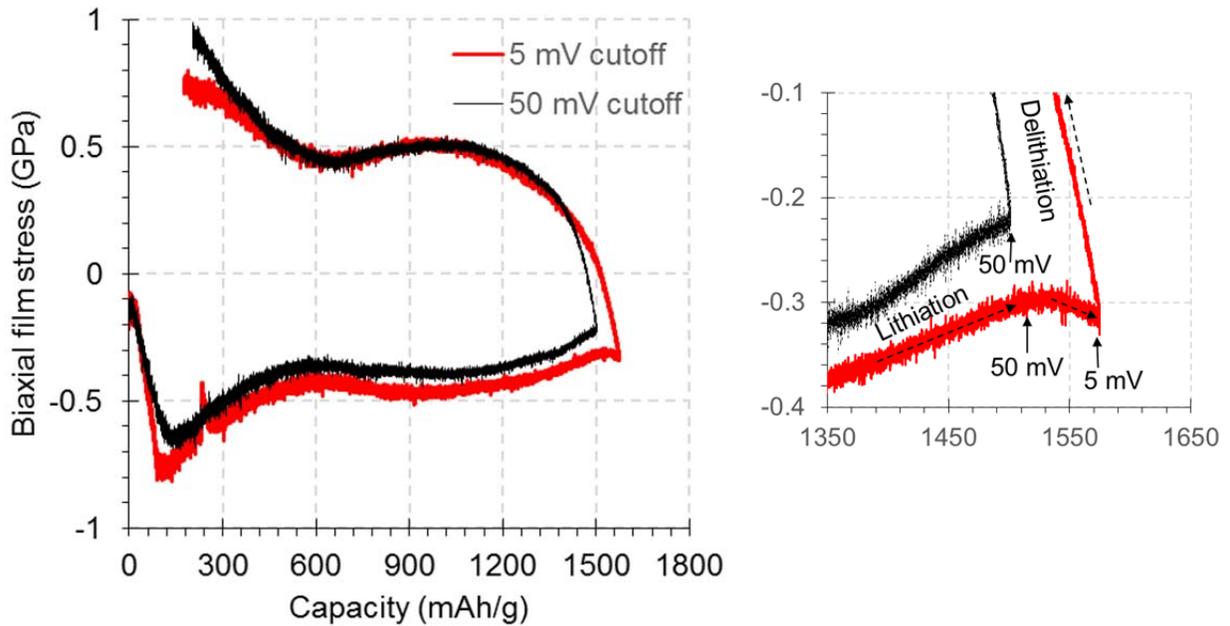

Figure 3: (a) Potential and (b) stress evolution as a function of capacity in Ge films during galvanostatic lithiation/delithiation at 5μA/cm$^2$ (which corresponds to C/17.5 rate in the current experiments). The two curves in the figures correspond to data from two different samples. The red curve represent the sample with a cut off potential of 5 mV vs Li/Li$^+$ and the black represents the sample with a cut off potential of 50 mV vs Li/Li$^+$. The inset shows that the stress versus capacity curve shows a distinct feature when lithiated below 50 mV corresponding to a phase change from amorphous Ge to crystalline Li$_{15}$Ge$_4$ phase.

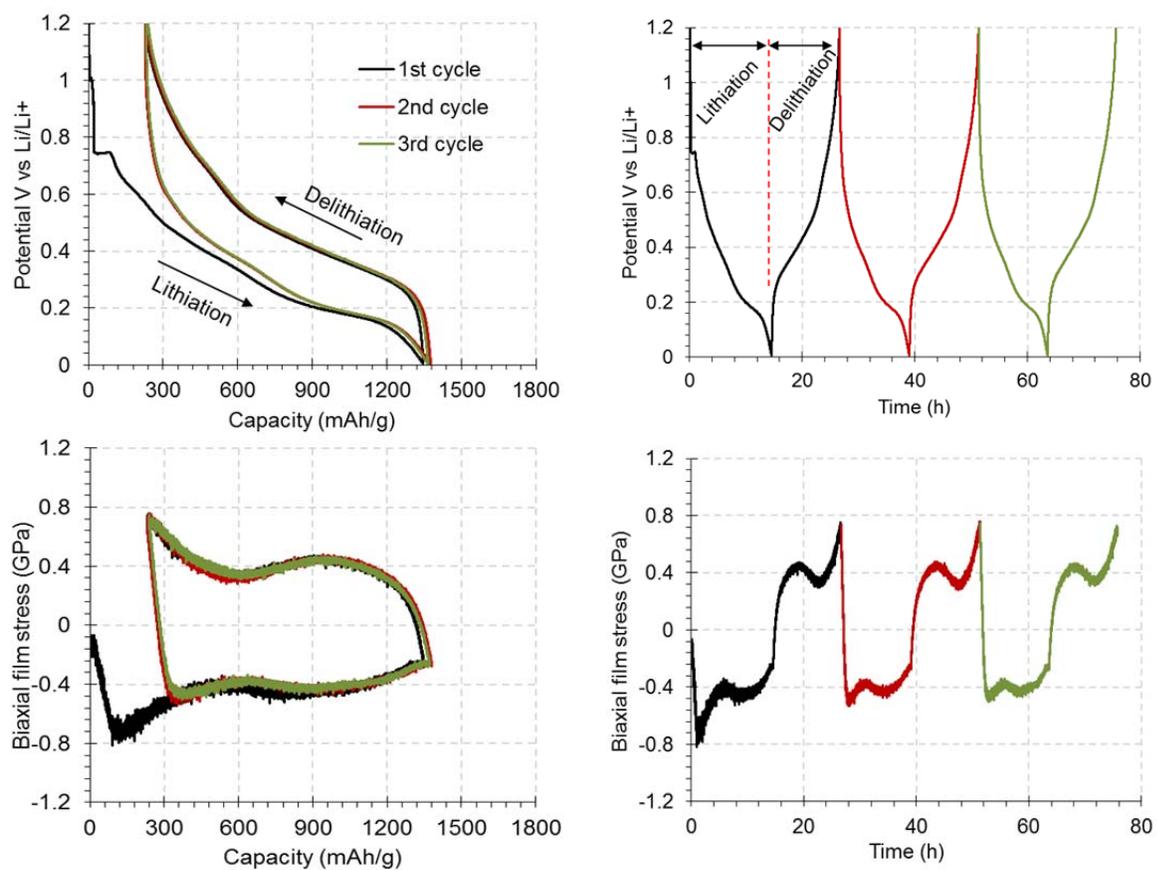

Figure 4: Potential and stress evolution in lithiated Ge films as a function of (a,c) capacity and (b,d) time, respectively under galvanostatic cycling at C/17.5 rate.

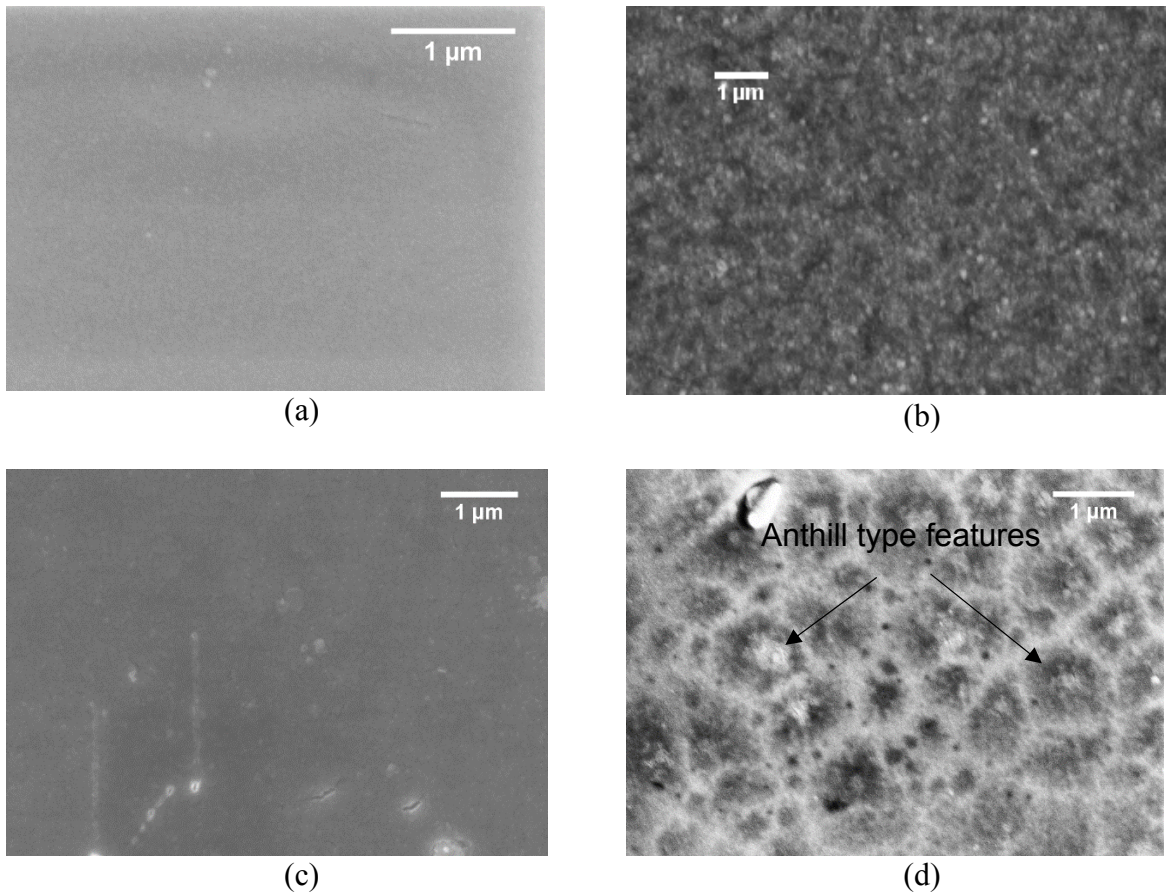

Figure 5: SEM images of (a) as prepared Ge thin film sample and (b) the film after one cycle of lithiation/delithiation (c) the film after 3 cycles, and (d) the film after several cycles. No cracks are visible after 3 to 6 cycles; however, the surface morphology of the film changed when subjected to more than 3 cycles, developing features that resemble anthills in (d).

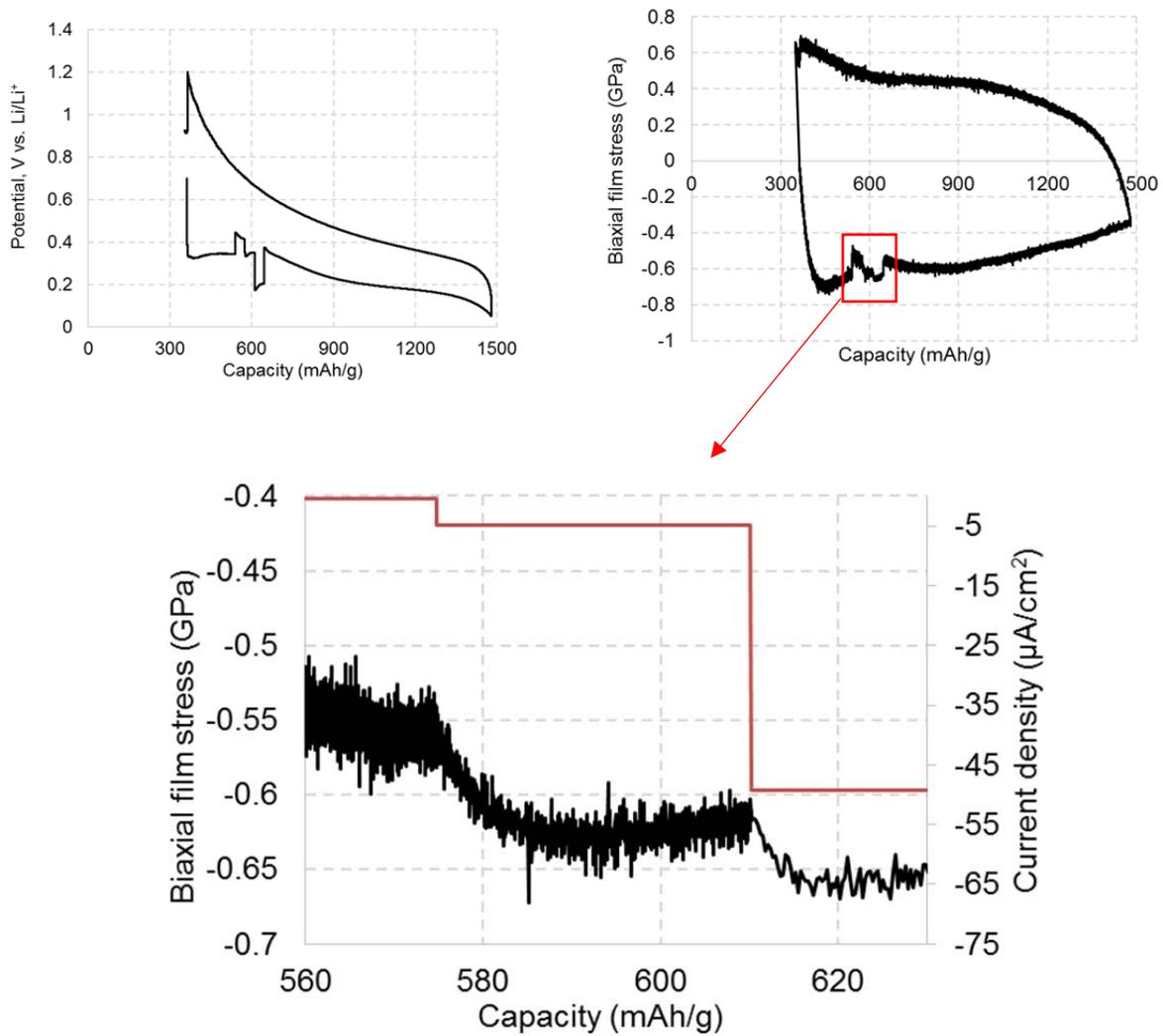

Figure 6: (a) and (b) show potential and stress evolution, respectively, of Ge film during a galvanostatic lithiation/delithiation experiment where the current density was varied at a particular state of charge to study the rate sensitivity of lithiated Ge film. The inset (c) shows the detailed stress response to change in current density.

Table 1 Material properties used in the analysis.

| Material | Young's Modulus (GPa) | Poisson's Ratio | Remarks |
| --- | --- | --- | --- |
| Lithiated Germanium (200 mAh/g) | 29 | 0.26 (assumed) | Obtained from the slope of stress-strain (capacity) at the beginning of $2^{nd}$ cycle lithiation. Procedure is similar to Sethuraman et al.[26] |
| Fused Silica | 71 | 0.16 | Ref. [16] |

**Figure captions**

Figure1: (a) Schematic of experimental setup and electrochemical cell; the inset shows details of thin film layers, and (b) shows the definition of various parameters used in Eq.1 and Eq.2 to measure curvature and stress. (c) Comparison of Raman spectra of sputter-deposited Ge thin films and polycrystalline (or Bulk) Ge. The broad peak of sputter-deposited film indicates that it is amorphous.

Figure 2: First three cycles from a cyclic voltammogram on a thin-film Ge electrode in 1M $LiPF_6$ in EC:DEC:DMC (1:1:1, wt.%) at a scan rate of 0.5 mV/s.

Figure 3: (a) Potential and (b) stress evolution as a function of capacity in Ge films during galvanostatic lithiation/delithiation at $5\mu A/cm^2$ (which corresponds to C/17.5 rate in the current experiments). The two curves in the figures correspond to data from two different samples. The red curve represent the sample with a cut off potential of 5 mV vs $Li/Li^+$ and the black represents the sample with a cut off potential of 50 mV vs $Li/Li^+$. The inset shows that the stress versus capacity curve shows a distinct feature when lithiated below 50 mV corresponding to a phase change from amorphous Ge to crystalline $Li_{15}Ge_4$ phase.

Figure 4: Potential and stress evolution in lithiated Ge films as a function of (a,c) capacity and (b,d) time, respectively under galvanostatic cycling at C/17.5 rate.

Figure 5: SEM images of (a) as prepared Ge thin film sample and (b) the film after one cycle of lithiation/delithiation (c) the film after 3 cycles, and (d) the film after several cycles. No cracks are visible after 3 to 6 cycles; however, the surface morphology of the film changed when subjected to more than 3 cycles, developing features that resemble anthills in (d).

Figure 6: (a) and (b) show potential and stress evolution, respectively, of Ge film during a galvanostatic lithiation/delithiation experiment where the current density was varied at a particular state of charge to study the rate sensitivity of lithiated Ge film. The inset (c) shows the detailed stress response to change in current density.